\documentclass[12pt]{article}
\usepackage{epsfig}
\def\eqn{\begin{equation}}
\def\enn#1{\label{#1} \end{equation}}
\def\eq{\[}
\def\en{\]}

\def\e#1{eq. (\ref{#1})}

\def\f#1{Figure {\ref{#1}}}
\def\t#1{Table \ref{#1}}
\def\hs#1{\hspace{#1}}

\begin{document} {
\setlength{\baselineskip}{12pt}

\pagestyle{plain}
\title{Constraint-defined Manifolds: \\
a Legacy Code Approach \\
to Low-dimensional Computation}

\author{
C. William Gear$^{1,2}$ and
Ioannis G. Kevrekidis$^{1}$\\
$^1$Department of Chemical Engineering; Princeton University,\\
$^2$NEC Research Institute (retired); \\
Princeton, NJ 08544, USA}

\maketitle

\begin{abstract}

If the dynamics of an evolutionary differential equation system
possess a low-dimensional, attracting, slow manifold, there are many advantages to
using this manifold to perform computations for long term dynamics,
locating features such as stationary points, limit cycles, or bifurcations.
Approximating the slow manifold, however, may be computationally as challenging as
the original problem.
If the system is defined by a legacy simulation code
or a microscopic simulator, it may be impossible to perform the
manipulations needed to directly approximate the slow manifold.
In this paper we demonstrate that with the knowledge only of a set of
``slow'' variables that can be used to {\it parameterize} the slow
manifold, we can conveniently compute, using a legacy simulator,
on a nearby manifold.
Forward and reverse integration, as well as the location of
fixed points are illustrated for a discretization of the Chafee-Infante
PDE for parameter values for which an Inertial Manifold is known to
exist, and can be used to validate the computational results.

\end{abstract}


\section{Introduction}

Certain dissipative evolutionary equations possess low-dimensional,
attracting invariant manifolds which govern their long term dynamics.
Such a manifold is readily apparent for a system given in the singularly perturbed form:
\eq
dy/dt = y' = f(y,z), \hs{0.2in} f \in \Re^{n+m} \mapsto \Re^n
\en
\eqn
dz/dt = z' = g(y,z)/\epsilon, \hs{0.2in}  g \in \Re^{n+m} \mapsto \Re^m
\enn{spde}
where $f$ and $g$ are such that, for small positive $\epsilon$,
$z$ is rapidly attracted to the region $z' = {\rm O}(1)$ and $\partial
g/\partial z$ is non-singular.
Since $z = z_0(y) + {\rm O}(\epsilon)$ where
$z_0(y)$ is the solution of $g(y,z_0(y)) = 0$, the slow manifold is
given by the solution of
\eqn
y' = f(y,z_0(y)) + {\rm O}(\epsilon)
\enn{man1}
so that we can ``easily'' compute an approximation to it.
The one-dimensional slow manifold is parameterized here by $y$
(other parameterizations are also possible).

In more complicated cases, approximations to the slow manifold may not be
so apparent; yet within such manifolds the system dynamics can still be described by a
lower-order differential equation - the {\em reduced system}.
Methods for approximating such manifolds have been the subject of
intense research in communities ranging from reactive flow modeling
(e.g. \cite{MaasPope,LamGoussis,GorbanCarlin}) to inertial manifolds
for dissipative PDEs (e.g. \cite{Constantin,Temam,JKT}).
If we are able to somehow constrain the dynamics to a slow manifold,
stable numerical integration could be performed with larger
stepsizes than would be possible in the original system.
Furthermore, many global properties of the original system are (approximately)
inherited by the reduced system; these include stationary points, limit cycles,
and bifurcations and may be computable more easily on the slow manifold.
Unfortunately, approximating the slow manifold may be as computationally challenging as the
original problem.

In our work we seek an approximation to such a manifold that is (a) simple to
obtain {\it on the fly} during numerical computations, and (b) only requires
evaluations of time derivatives of the state, such as would be available
from a legacy code.
Our starting point is the assumption that, given a basis for the full set of
variables in the problem, a subset of this basis can be used
to parameterize the slow manifold and our approximation of it,
as $y$ did in our example above.
In some applications, such as when the full system is described by a
microscopic simulation, the subset used to parameterize the slow
manifold might be called ``macroscopic observables"; such observables could
be the pressure field in kinetic theory based flow simulation,
or a concentration field in the
kinetic Monte Carlo simulation of a chemical reaction.

We may start with a finite-dimensional system (an ODE) or an
infinite-dimensional system (a PDE, for example).
In the latter
case we will have to introduce a finite-dimensional approximation
before commencing computation (in effect, the ``method of lines'').
Suppose that the system is represented by (approximated
by) the system \eqn u_t = L(u) \enn{pde} where $u = \{u_i\},  i = 1,
\cdots, N$.
For ease of presentation, let us assume that the
equation has already been transformed to a suitable basis so that $U = \{U_i =
u_i, \hs{0.1in} i = 1, \cdots, n\}$ parameterizes the slow manifold.
In some sense we are
assuming that the remaining variables, $V = \{V_i =
u_{n+i},\hs{0.1in} i = 1, \cdots, N-n = m\}$, are the ``fast''
ones that are quickly ``slaved'' to $U$; we will return to this assumption.
The split system is \eqn
\frac{dU}{dt} = L_1(U,V) \enn{1a} \eqn \frac{dV}{dt} = L_2(U,V).
\enn{1b}
We now view the solutions of the system as families of curves
in the $U-V$ state space, as illustrated in \f{F1} - although such
figures are potentially misleading because we have to remember
that each axis represents a multi-dimensional space.
\begin{figure}[p]
\centerline{\psfig{figure=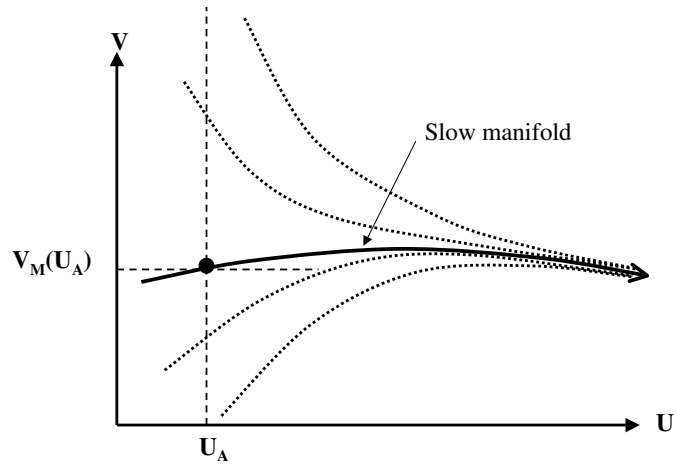,height=3.5in,height=3.5in}}
\vspace{0.1in} \caption{Family of Solutions and Slow
Manifold Schematic}\label{F1}
\end{figure}
The essential feature of the figure is that one member of the family of solutions
is a ``slow'' manifold {\it with no high-curvature region}, while other members of the
family of solutions approach this slow manifold relatively rapidly.

Since the slow manifold, $M$, can be parameterized by the slow
variables, $U$, points on $M$, $(U,V_M(U))$, must be uniquely determined by $U$
- that is, the curve cannot ``fold'' in the region of interest.
If we had a scheme for approximating the value of
$V = V_M(U)$ for each $U$ (as we did in the singularly perturbed
example above) we could, for example, apply a numerical
integration method to just the $U$ variables, {\it computing the
equivalent values of $V_M(U)$ only as needed by the integration scheme}.
This is the main point of our approach: one does not compute the entire
manifold {\it a priori}, but only computes it pointwise, ``on demand" as required
by the low-dimensional integration code (or by algorithms performing
other numerical tasks, such as fixed point computation).

From the assumptions, we suspect that an approximation to the slow
manifold can be found by computing the value of $V = V_A(U)$ (as
shown in \f{F4}) for which the time derivatives of the $V$
components are zero.
Here we will compute on this ``steady manifold"; for the appropriate
basis choice this steady AIM is not too far from the slow manifold.
Better approximations, based on higher order expansions of singularly
perturbed equations, can also be used in a legacy code context, and
will be the subject of further work.

\begin{figure}[p]
\centerline{\psfig{figure=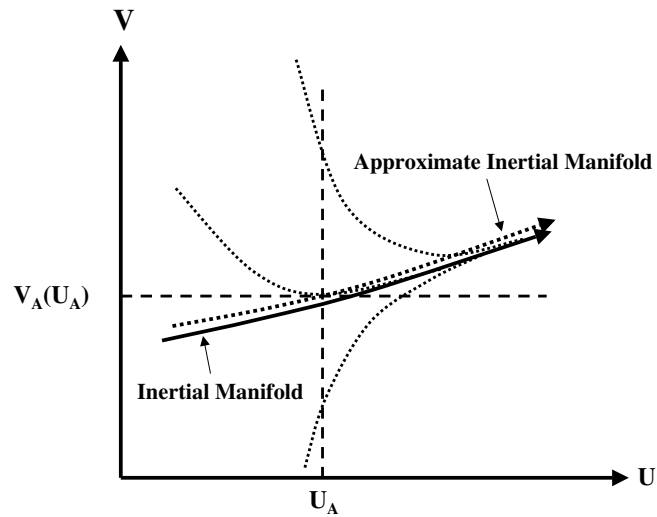,height=3.5in,height=3.5in}}
\vspace{0.1in}
\caption{Approximate Inertial Manifold Definition}
\label{F4}
\end{figure}

We are particularly interested in the case where we do not have
``access'' to the differential equations directly because, for
example, we have a legacy simulation code, or the system is the unknown
closure of a microscopic simulation  model (kinetic Monte Carlo,
molecular dynamics).
Then, the only computational possibility we have is to integrate the full
system for a short time in what we call an {\em inner} integration step.
In this case, we can define an AIM by
requiring that the {\em chord} of the inner step has zero change in
the $V$ components, that is, it is ``horizontal.''
This can be computed iteratively by
performing an inner integration over a small step of length $h$ and
then {\it projecting} the solution back to the specified value of $U$ (by
simply resetting the values of $U$ to their values at
the start of the step) and repeating.
If $h$ is small enough, this
iteration will converge if the solution family in the neighborhood of
$(U, V_A(U))$ is locally attracted to the solution that passes through
$(U,V_A(U))$.
This can be seen by noting that one step of the process
performs the mapping
\eqn
V_{m+1} = V_m + gL_2(U,V_m) + {\rm O}(h^2)
\enn{it1}
so we have
\eq
V_{m+1} - V_m = (I + h\frac{\partial L_1}{\partial V})(V_m - V_{m-1})
+ {\rm O}(h^2)
\en
which implies convergence for small enough $h$ if the eigenvalues of
$\partial L_1/\partial V$ are in the negative half plane.
This property can form the basis of alternative algorithms to approximate
the ``steady AIM:" matrix free fixed point algorithms, like the Recursive
Projection Method or GMRES \cite{Shroff,Kelley}
can be applied to accelerate the computation of
the fixed point of \e{it1}.

In the following sections we will illustrate the use of this
technique on the Chafee-Infante reaction-diffusion equation \eqn \frac{\partial
u}{\partial t} = \nu \frac{\partial^2 u}{\partial x^2} + u - u^3
\enn{rde} with $\nu = (2.5)^{-2}$ and $u(0,t) = u(\pi,t) = 0$.
This is known to possess an inertial manifold of dimension two (in effect,
the two-dimensional unstable manifold of the origin, and its closure).
Although we know the differential equations in this example, we are not going to make
explicit use of the knowledge in our computational method.
We will only use it as if we were given a legacy code for evaluating
time derivatives.

We first discretize the equations in space.
Since we are not interested in the issue of the best
spatial discretization, we use simple finite difference
methods over $N$ equally spaced points, so that the variables are
$u_i(t) = u(x_i,t), i = 1, \cdots, N$, where $x_i = \pi i/(N+1)$.
These variables are chosen for convenience in the calculation.
The resulting ODEs are the usual: \eqn \frac{du_i}{dt} =
\frac{\nu}{\Delta X^2}(u_{i-1} - 2u_i + u_{i+1}) +u_i -
u^3_i,\hs{0.2in} i = 1, \cdots, N \enn{semid} where $u_0 = u_{N+1}
= 0$.
If we had a legacy code or a microscopic model, the
variables would be the ones that happened to come with the code or
model.
In this example, no subset of the $\{u_i\}$ variables is suitable for
defining the AIM (since the slow manifold varies rapidly as a function
of each $u_i$) so we will use an ``observation basis" in which a linear
combination of the variables will parameterize the AIM.
In this
case, we can use a basis formed by $\sin(mx), m = 1, \cdots, N$.
(These are the unnormalized eigenvectors of \e{semid} when $u=0$.)
The modified variables are
\eqn
a_i = \sum_{j=1}^N\phi_{ij} u_j \hs{0.2in}
\enn{basis}
where
$\phi_i = \{\phi_{i*}\}$, is the basis given by $\phi_{ij} = \sin(ix_j)$.
The first two $a_i$ can parameterize the slow manifold, and it is not necessary to
calculate the rest.

We now present a technique for approximating $d(a_1, a_2)/dt$ {\it on
the slow manifold} given $(a_1,a_2)$.
This approximation can then be used to implement
time integration, stability analysis, or other numerical
procedures on the system {\it constrained to the AIM}.
The general method consists of
\begin{enumerate}
\item Start with a prescribed value of $(a_1,a_2)$.
\item Compute the values of $\{u_i\}$ such that \e{basis} is satisfied
and the local derivative of the full ($\{u_i\}$ system) is
``horizontal'' in the other components of the basis (in this
example, $a_3, \cdots, a_N$). This can be done in a number of ways:
\begin{enumerate}
\item Use \e{semid} and \e{basis} to compute $da_i/dt, 3 \le i = \le
N$ and then solve for the values of $a_j$ that makes these zero using
Newton iteration.  This can be done directly when the equations are available (it is
done in the example illustrated here), or can be done through matrix-free
based contraction mappings if the equations are not explicitly available.
\item If we only have a legacy code or a
microscopic simulator of the full system, use iteration \e{it1} repeatedly to
find the values of $a_j, 3 \le j \le N$ such that the chords of
those $a_j$ are zero.
\item Conceptually (since this is not practical for a legacy code) one could
    implement a Lagrange multiplier, evolving the dynamics while
    constraining the projection of the solution on $(a_1,a_2)$.
    This is reminiscent of techniques like SHAKE used to ``prepare"
    molecular dynamics simulations \cite{SHAKE}. The approach
    described immediately above is a way of effectively implementing
    what amounts to such a Lagrange multiplier constrained integration
    to a legacy simulator.
\end{enumerate}
\item Compute the derivatives (or chord slope) of the full ($\{u_i\}$)
system from the given values of $a_1, a_2$ and the now computed $a_j,
3 \le j \le N$
(actually they have probably been computed in the previous step).
\item Compute the ``$(a_1,a_2)$" components of the derivative by
applying \e{basis} to the $\{u_i\}$ derivatives.
These are the
approximations to the time derivatives of $(a_1,a_2)$ {\it on the steady AIM}.
\end{enumerate}

In the next section we will use this technique to integrate \e{semid}
both forward and backward in time on our two-dimensional steady AIM,
and compare it with the
integration of the full system and in the subsequent section we will
use it to compute the steady states directly by performing a Newton
iteration on the two-dimensional steady AIM.

\section{Integration on an AIM of the Reaction-Diffusion Equation}

In \cite{cwgygk} we introduced {\it projective integration} which uses
computation of the chord slope obtained by integration of a legacy
code or of a microscopic model in place of derivatives for
performing large {\em projective} integration steps on the
slow components.
If we were working with legacy codes or
microscopic simulators, we would use that technique in our ``on manifold"
integrations.
However, we have chosen an example for which we
know the equations of the detailed system, so that we can compare
the ``true'' integration of the system with the approximation on
the steady AIM we have defined.

We integrated the full system \e{semid} using the
automatic Runge-Kutta method with the Dormand-Prince pair of
formulae known as RK45 and available in MATLAB as {\bf ode45}.
We also
used the same method to integrate just $(a_1,a_2)$ {\it on the AIM},
using the technique described in the previous section to
approximate the derivatives of $(a_1,a_2)$.
It is possibly better to view this as the (approximate) integration {\it of the
projection} on the $(a_1,a_2)$ ``observables plane" of the true dynamics.
The results for $(a_1,a_2)$ are shown in \f{figsh}.
The integration was started
from six different points near the origin in the $(a_1,a_2)$ plane
(which is an unstable steady state).
All but one approach the
stable steady state at approximately (1.12,0) but the one that
starts at (0,0.05) stays on the invariant submanifold $a_1 = 0$ and moves to the saddle
point at about (0,0.7).
Since the origin is on the inertial
manifold of \e{semid}, the starting points are also very close to it
so that the RK45 integration of the full system
gives a good approximation of the solutions on the inertial manifold and
provides a picture of the manifold itself.
In \f{figsh},
the solid line is the RK45 solution of the full system, while the
dashed line is the integration on our steady AIM.
\begin{figure}[p]
\centerline{\psfig{figure=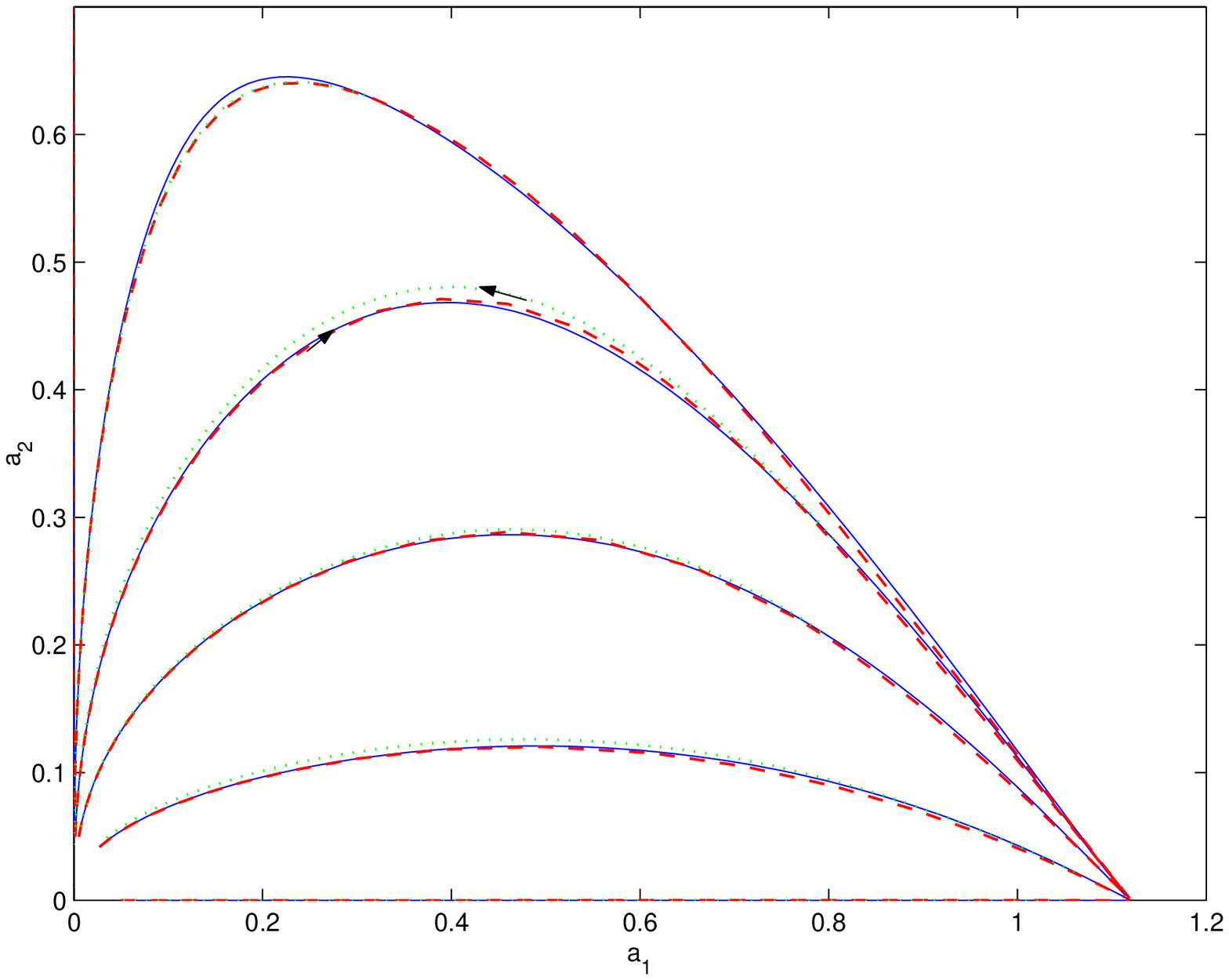,height=3.5in,}}
\vspace{0.1in} \caption{Trajectories in $(a_1,a_2)$-plane. Solid:
Inertial manifold, dashed - AIM forward, dotted - AIM reverse.
}\label{figsh}
\end{figure}

The full system in \e{semid} is rapidly damped in its fastest components, and
so it would not be feasible to numerically integrate it {\it in the reverse time
direction}.
However, the $(a_1,a_2)$ differential system on the
AIM does not have these fast components, so it can be integrated
``backwards.''
The dotted arcs in \f{figsh} are the results of a
reverse integration in the $(a_1,a_2)$-plane starting from a point
on the RK45 solution of the full system shortly before the
stationary point is reached (one can't start too close to the
stationary point because the trajectory chosen would be too
sensitive to small perturbations).

As we can see in \f{figsh}, the forward and reverse solutions on
the AIM are fairly good approximations to the components of the
``true" solution on the IM.
We do not expect the other
components, $a_i, i \ge 3$ to be good approximations.
This is shown in \f{fsh13} which shows the values of $a_3$ plotted against
$a_1$ for each of the trajectories.
(The reverse integration
trajectories are almost indistinguishable from the forward trajectories
for the $(a_1,a_2)$ integration.)
\begin{figure}[p]
\centerline{\psfig{figure=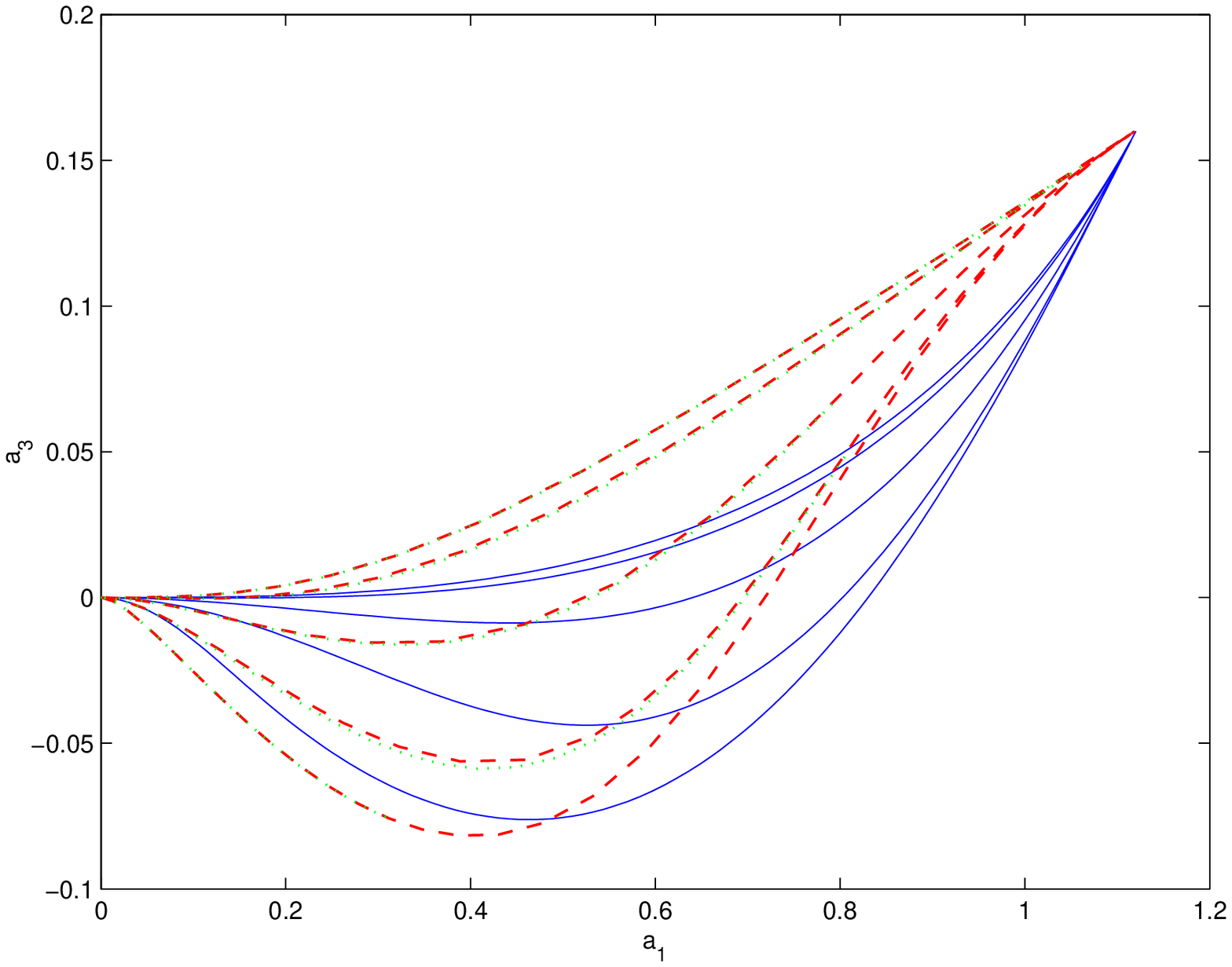,height=3.5in,height=3.5in}}
\vspace{0.1in} \caption{$(a_1,a_3)$ trajectories. Solid: Inertial
manifold, dashed - AIM forward, dotted - AIM
reverse.}\label{fsh13}
\end{figure}

\section{Steady State Computation on the AIM}

The procedure described in Section 1 computes $da/dt = p(a)$ where $a
= (a_1,a_2)$ is the parameterization of the slow manifold.
Using any standard techniques we can look for zeros of $p(a)$ to identify
stationary states of the system.
Since the dimension is low, we can
use Newton's method, computing the approximate partial derivatives by
finite differencing.
Many better methods exist, but our purpose here
is simply to show that the reduced system can be used directly in any
conventional numerical process.

\t{T1} shows the sequence of iterates for Newton's method
starting at three different point in the $(a_1,a_2)$-plane and
iterating until changes were less than $10^{-5}$ in the L1 norm.
The
eigenvalues of $J = \partial p/\partial a$ as the iteration proceeds
are also shown.
For comparison, the two leading eigenvalues of the
full system \e{semid} at steady state for each of the three cases  are
(0.8403,0.3648), (0.2204,-0.7118), and (-1.4491,-1.5392) respectively.
As can be seen, the three stationary points, (0,0), (0,0.7056), and (1.1206,0)
are source, saddle, and stable (sink), respectively.
Note that the stationary
states on the steady AIM are necessarily on the slow manifold since all of
the derivatives are zero at these points.

\begin{table}
\begin{center}
\caption{Sequence of
Newton Iterates}
\vspace{0.1in}
\begin{tabular}{|r|r|r|r|} \hline
\multicolumn{4}{|c|}{Case 1}\\ \hline
\multicolumn{1}{|c|}{$a_1$} &\multicolumn{1}{|c|}{$a_2$} &
\multicolumn{1}{|c|}{$\lambda_1$} &
\multicolumn{1}{|c|}{$\lambda_2$} \\ \hline
    0.2000 &   0.2000 &   0.7171 &   0.1971\\
   -0.0838 &  -0.1946 &   0.7737 &   0.2684\\
    0.0159 &   0.0551 &   0.8351 &   0.3571\\
   -0.0002 &  -0.0009 &   0.8403 &   0.3648\\
    0.0000 &   0.0000 &   0.8403 &   0.3648\\ \hline
\multicolumn{4}{|c|}{Case 2}\\ \hline
\multicolumn{1}{|c|}{$a_1$} &\multicolumn{1}{|c|}{$a_2$} &
\multicolumn{1}{|c|}{$\lambda_1$} &
\multicolumn{1}{|c|}{$\lambda_2$} \\ \hline
    0.1000 &   0.7500 &   0.1666 &  -0.8904\\
   -0.0405 &   0.7271 &   0.2069 &  -0.7902\\
    0.0063 &   0.7087 &   0.2359 &  -0.7293\\
   -0.0001 &   0.7057 &   0.2405 &  -0.7203\\
    0.0000 &   0.7056 &   0.2407 &  -0.7200\\ \hline
\multicolumn{4}{|c|}{Case 3}\\ \hline
\multicolumn{1}{|c|}{$a_1$} &\multicolumn{1}{|c|}{$a_2$} &
\multicolumn{1}{|c|}{$\lambda_1$} &
\multicolumn{1}{|c|}{$\lambda_2$} \\ \hline
  1.0000      &    0.1000  &     -0.8687  &  -1.5201\\
  1.1707      &   -0.0462  &     -2.0297  &  -1.6832\\
  1.1255      &   -0.0042  &     -1.6847  &  -1.6599\\
  1.1206      &   -0.0000  &     -1.6529  &  -1.6529\\
  1.1206      &   -0.0000  &     -1.6528  &  -1.6526\\ \hline
\end{tabular}
\label{T1}
\end{center}
\end{table}

\section{Discussion}

We have demonstrated that it is possible to perform low-dimensional
(macroscopic) computations on an AIM (more precisely,
on {\it observations} of an AIM)
based on choosing a suitable
parameterization of the low dimensional slow manifold.
That parameterization must be chosen so that the slow manifold does not
``fold over it".
It must also be chosen so that the induced AIM is
reasonably close to the true slow manifold.
In the example we
discussed, the first two eigenvectors of the linearization of the
problem at a
particular solution value (the origin) were chosen to parameterize
the manifold, since at that solution value they are tangent to the
true slow manifold.
As long as
the solution does not stray too far from that region (compared to
the non-linearities present) these directions provide a reasonable
parameterization to the slow manifold elsewhere.
The steady AIM for these
two variables is illustrated in \f{man123} which plots $a_3$
against $a_1$ and $a_2$ on the AIM.
This AIM is a reasonable
approximation of the slow manifold, (which is shown is Figure 1
of \cite{yannis}).
It is clear from this figure that choosing,
say, $a_2$ and $a_3$ to characterize the slow manifold would have
been bad since $a_1(a_2,a_3)$ is multivalued for some $a_3$. (See
the line $a_2 = 1, a_3 = -0.1$ that is indicated in the
figure.  It intersects the AIM twice within the region plotted.)
This is true even near the origin, where this AIM is a good
approximation to the inertial manifold.

The focus of this work was on the use of a legacy simulator
to approximate the slow manifold on the fly, as dictated by
the needs of numerical analysis tools employed for computations on it.
The local nature of the approximation should be contrasted to
``off line" algorithms that attempt to approximate the entire
manifold first (see the extensive discussion in \cite{Gorban}
as well as \cite{JonesTiti}).
Here we pursued the simplest approximate manifold one can find
by constraining a legacy code.
An important issue that was only tangentially mentioned here was
the selection of good basis functions (or macroscopic observables
in the case of atomistic inner simulators) that parameterize the
manifold; statistical data analysis techniques have an important
role to play in this.
Better algorithms, resulting
from the implementation of higher order approximations to the
slow manifold (requiring a higher order derivative to vanish)
in a legacy code context are currently being explored.

{\bf Acknowledgements} This work was partially supported by
an NSF/ITR grant and by the AFOSR Dynamics and Control (Dr. B. King).

\begin{figure}[p]
\centerline{\psfig{figure=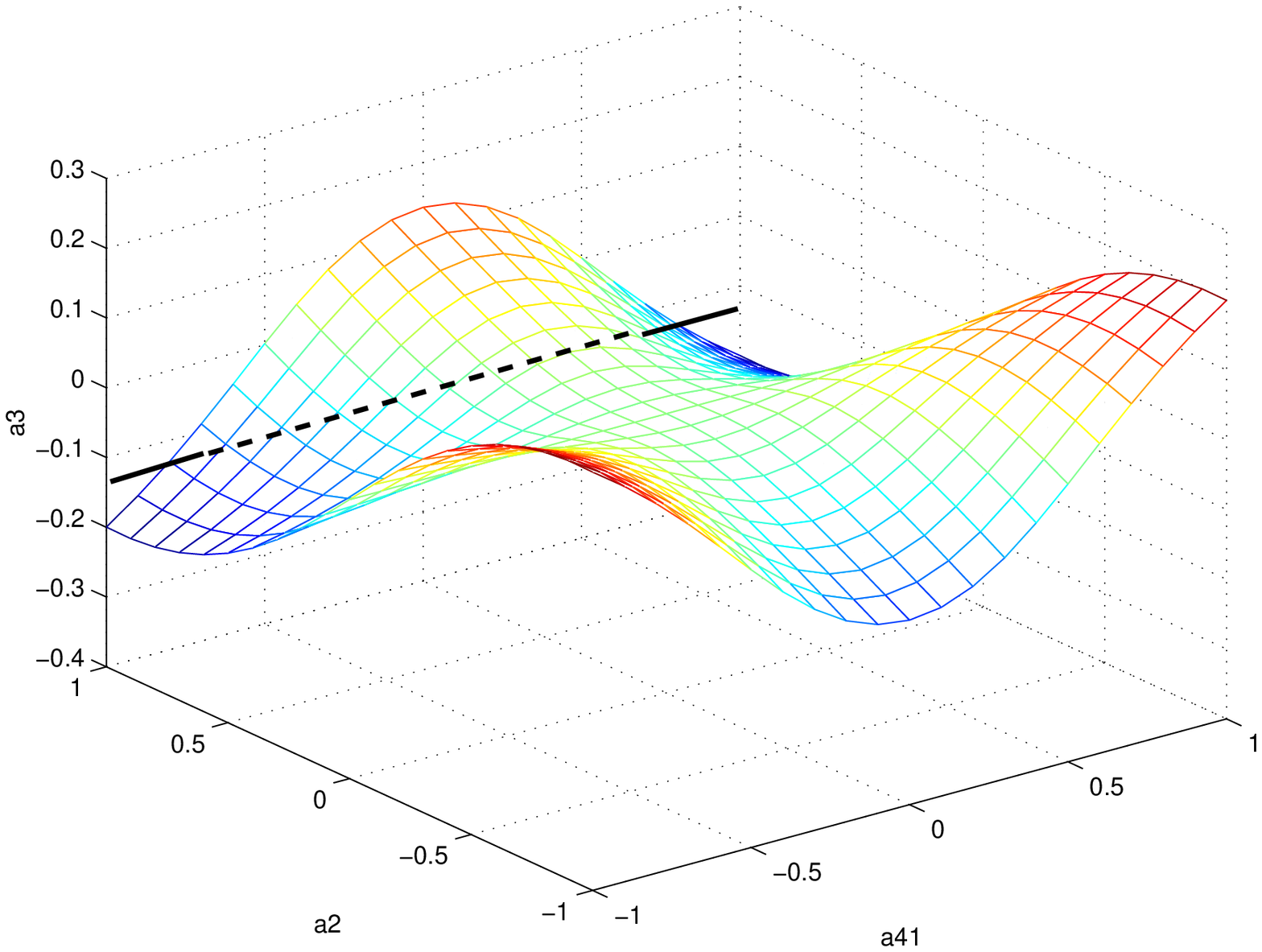,height=3.5in,height=3.5in}}
\vspace{0.1in} \caption{$a_3$ as a function of $a_1$ and $a_2$ on the AIM.}\label{man123}
\end{figure}

}
\newpage   
\end{document}